\documentclass[letter]{aa} 

\def\Msol      {$\hbox{M}_\odot$}

\def\THCO      {$^{13}$CO}

\def\water     {H$_2$O}

\def\sioI      {$^{28}$SiO}
\def\sioII     {$^{29}$SiO}
\def\sioIII    {$^{30}$SiO}

\def\siI      {$^{28}$Si}
\def\siII     {$^{29}$Si}
\def\siIII    {$^{30}$Si}

\def\dsiII     {$\delta^{29}$Si}
\def\dsiIII    {$\delta^{30}$Si}

\def\kms       {km~s$^{-1}$}

\def\vlsr      {$V_{\rm LSR}$}

\def\Eup       {$E_{\rm up}/k$}

\usepackage{graphicx}
\usepackage{lscape}
\usepackage{natbib}
\usepackage{txfonts}
\usepackage{multirow}
\usepackage{longtable}
\usepackage{rotating}
\begin{document}
   \title{Silicon isotopic abundance toward evolved stars and its application for presolar grains
   \thanks{This publication is based on data acquired with the Atacama Pathfinder Experiment (APEX). APEX is a collaboration between the Max-Planck-Institut f\"ur Radioastronomie, the European Southern Observatory, and the Onsala Space Observatory. {\it Herschel} is an ESA space observatory with science instruments provided by European-led Principal Investigator consortia and with important participation from NASA.}
   }
   \author{T.-C. Peng\inst{1}
         \and E. M. L. Humphreys\inst{1}
         \and L. Testi\inst{1,2,3}
         \and A. Baudry\inst{4,5}
         \and M. Wittkowski\inst{1}
         \and M. G. Rawlings\inst{6}
         \and I. de Gregorio-Monsalvo \inst{1,8}
         \and W. Vlemmings \inst{7}
         \and L.-A. Nyman \inst{8}
         \and M. D. Gray \inst{9}
         \and C. de Breuck \inst{1}
          }
   \institute{ESO Garching, Karl-Schwarzschild Str. 2, D-85748 Garching, Germany \\
        \email{tpeng@eso.org}      
        \and Excellence Cluster Universe, Boltzmannstr. 2, D-85748 Garching, Germany
        \and INAF-Osservatorio Astrofisico di Arcetri, Largo E. Fermi 5, I-50125 Firenze, Italy 
        \and Univ. Bordeaux, LAB, UMR 5804, F-33270, Floirac, France
        \and CNRS, LAB, UMR 5804, F-33270, Floirac, France   
        \and National Radio Astronomy Observatory, 520 Edgemont Road, Charlottesville, VA 22903, USA
        \and Department of Earth and Space Sciences, Chalmers University of Technology, Onsala Space Observatory, SE-439 92 Onsala, Sweden
        \and Joint ALMA Observatory (JAO) and European Southern Observatory, Alonso de C\'ordova 3107, Vitacura, Santiago, Chile
        \and JBCA, Alan Turing Building, School of Physics and Astronomy, University of Manchester, Manchester M13 9PL, UK
             }

   \date{}

 
  \abstract
   {}
   {Galactic chemical evolution (GCE) is important for understanding the composition of the present-day interstellar medium (ISM) and of our solar system. In this paper, we aim to track the GCE by using the \siII/\siIII\ ratios in evolved stars and tentatively relate this to presolar grain composition.}
   {We used the APEX telescope to detect thermal SiO isotopologue emission toward four oxygen-rich M-type stars. Together with the data retrieved from the {{\it Herschel}} science archive and from the literature, we were able to obtain the \siII/\siIII\ ratios for a total of 15 evolved stars inferred from their optically thin \sioII\ and \sioIII\ emission. These stars cover a range of masses and ages, and because they do not significantly alter \siII/\siIII\ during their lifetimes, they provide excellent probes of the ISM metallicity (or \siII/\siIII\ ratio) as a function of time.}
   {The \siII/\siIII\ ratios inferred from the thermal SiO emission tend to be lower toward low-mass oxygen-rich stars (e.g., down to about unity for W Hya), and close to an interstellar or solar value of 1.5 for the higher-mass carbon star IRC+10216 and two red supergiants. There is a tentative correlation between the \siII/\siIII\ ratios and the mass-loss rates of evolved stars, where we take the mass-loss rate as a proxy for the initial stellar mass or current stellar age. This is consistent with the different abundance ratios found in presolar grains.  Before the formation of the Sun, the presolar grains indicate that the bulk of presolar grain already had \siII/\siIII\ ratios of about 1.5, which is also the ratio we found for the objects younger than the Sun, such as VY CMa and IRC+10216. However, we found that older objects (up to possibly 10 Gyr old) in our sample trace a previous, lower \siII/\siIII\ value of about 1. Material with this isotopic ratio is present in two subclasses of presolar grains, providing independent evidence of the lower ratio. Therefore, the \siII/\siIII\ ratio derived from the SiO emission of evolved stars is a useful diagnostic tool for the study of the GCE and presolar grains.}
%
   {}

   \keywords{ISM: abundances, ISM: molecules, Submillimeter: ISM, Stars: late-type}

\titlerunning{Silicon isotopic abundance toward evolved stars}
   \maketitle
%
\section{Introduction}

As the eighth most abundant element in the Universe, silicon plays an important role in understanding nucleosynthesis and Galactic chemical evolution (GCE). The main isotope \siI\ is mainly produced by early-generation massive stars that become Type II supernovae. The other two stable isotopes \siII\ and \siIII\ are mainly produced by O and Ne burning in massive stars or by slow neutron capture (the {\it s}-process) and by explosive burning in the final stages of stellar evolution, that is, the asymptotic giant branch (AGB) phase for low- and intermediate-mass stars and supernova explosions for high-mass stars \citep[see, e.g.,][]{Woosley1995,Timmes1996,Alexander1999}. 

In the thermally pulsing AGB (TP-AGB) phase, thermonuclear runaways are periodically caused by He burning in a thin shell between the H-He discontinuity and the electron-degenerate C-O core. This energy goes directly into heating the local area and raises the pressure, which initiates an expansion and a series of convective and mixing events \citep[][]{Herwig2005,Iben1983}. During the so-called third dredge-up, the products of He burning and the {\it s}-process elements are brought to the surface, e.g., $^{12}$C, which can lead to the formation of S- (C/O$\approx$1) or C-type (C/O$>$1) stars. In conjunction with dredge-ups, the Si-bearing molecules (e.g., SiC and SiO) formed in the stellar surface eventually condense onto dust grains or actually form dust grains. The silicon isotopic ratios will be preserved and go through the journey in the interstellar medium (ISM) until they are used again to form stars. 


AGB stars can produce almost all grains of interstellar dusts, and their dust production is one order of magnitude higher than that of supernovae in the Milky Way \citep[see, e.g.,][]{Dorschner1995,Gehrz1989}. It is generally believed that oxygen-rich M-type stars produce mainly silicate grains and carbon-rich stars mainly carbonaceous grains \citep{Gilman1969}. However, the actual situation may be more complicated and grain composition may change during the AGB phase \citep[][]{Lebzelter2006}. 



The measured \siII/\siIII\ ratios in the ISM are about 1.5 \citep{Wolff1980,Penzias1981}, very close to that of the solar system \citep{Anders1989,Asplund2009}. However, near-infrared SiO observations of \citet{Tsuji1994} showed that some evolved stars have \siII/\siIII\ ratios slightly below 1.5. Our new observations of SiO isotopologues in the radio domain with the APEX and {\it Herschel} telescopes confirm the low \siII/\siIII\ ratios for oxygen-rich M-type stars.





\begin{table}
\caption{Spectral parameters of the observed SiO isotopologue transitions}             
\label{table-1}      
\centering                          
\begin{tabular}{lrrrc}        
\hline\hline                 
Line         &  Frequency               & $E_{\rm up}/k $ & $\theta_{\rm MB}$ & Instrument \\    
             &  (MHz)                   & (K)             & (\arcsec) &               \\
       
\hline

\sioI\   $\upsilon$=0, $J$=6--5   & 260518.02   &  43.8 & 24.0 & APEX-1 \\
\sioII\  $\upsilon$=0, $J$=6--5   & 257255.22   &  43.3 & 24.3 & APEX-1  \\
\sioIII\ $\upsilon$=0, $J$=6--5   & 254216.66   &  42.7 & 24.5 & APEX-1  \\
\sioII\  $\upsilon$=0, $J$=7--6   & 300120.48   &  57.7 & 20.8 & APEX-2  \\
\sioIII\ $\upsilon$=0, $J$=7--6   & 296575.74   &  57.0 & 21.1 & APEX-2  \\
\sioII\  $\upsilon$=0, $J$=26--25 & 1112832.94  &  721.6 & 19.8 & HIFI \\
\sioIII\ $\upsilon$=0, $J$=26--25 & 1099711.49  &  713.1 & 19.6 & HIFI \\

\hline                                   
\end{tabular}

\tablefoot{$\theta_{\rm MB}$ is the FWHM beam width at the observed frequencies. The Kelvin-to-Jansky conversions are 39, 41, and 390 Jy K$^{-1}$ for the APEX-1, 2, and HIFI observations, respectively.  
}

\end{table}
   

   \begin{figure}
   \centering
   \includegraphics[angle=0,width=0.45\textwidth]{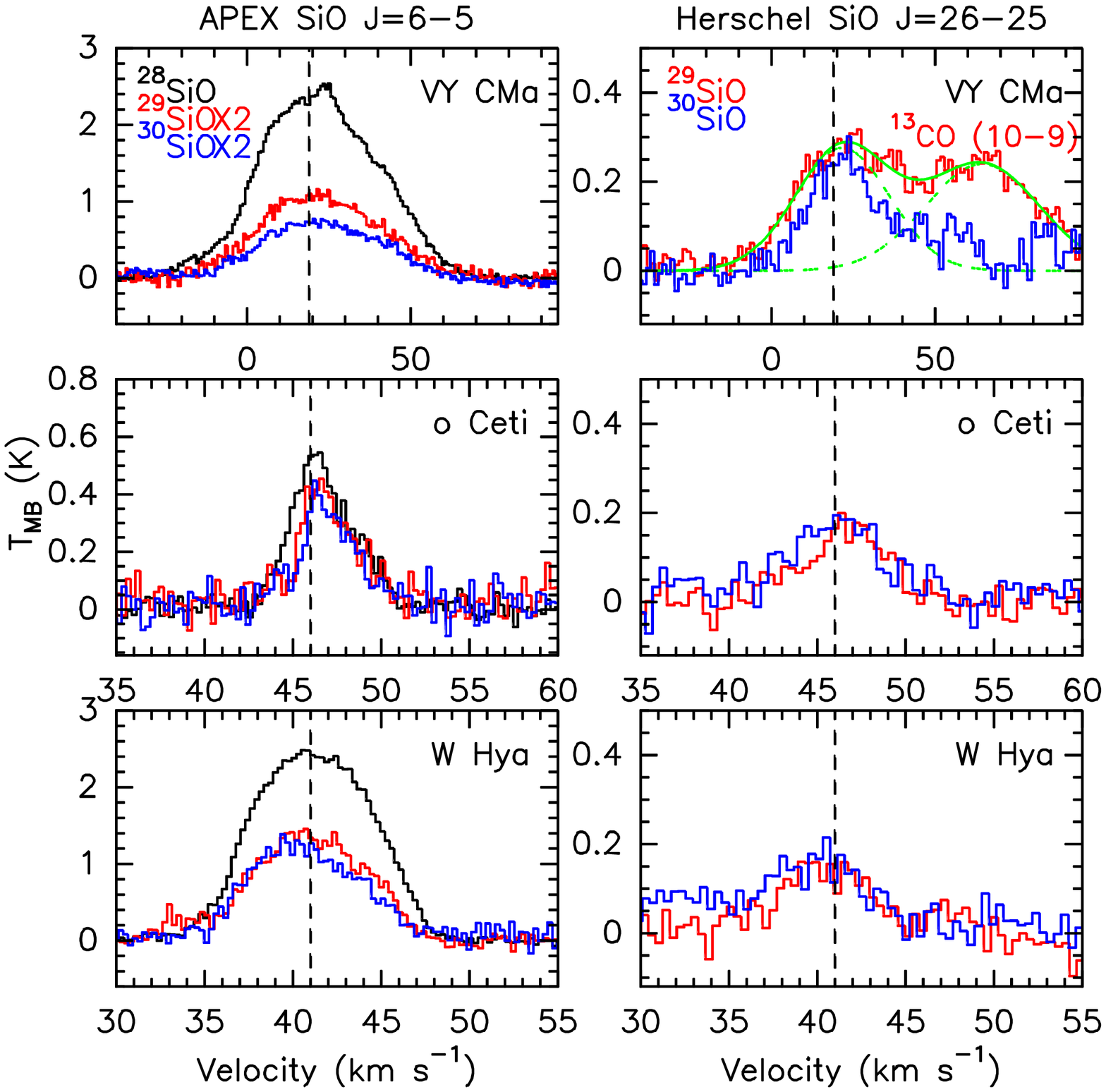}
   \caption{Left: APEX ground-vibrational \sioI\ (black), \sioII\ (red), and \sioIII\ (blue) $J$=6--5 spectra toward VY CMa, $o$ Ceti, and W Hya. The intensities of the \sioII\ and \sioIII\ lines were multiplied by two for clarity. Right: $Herschel$/HIFI ground-vibrational \sioII\ (red) and \sioIII\ (blue) $J$=26--25 spectra at around 1.1 THz toward the same sources. The \sioII\ emission of VY CMa is blended by the \THCO\ $J$=10--9 line from the other sideband. The dashed lines indicate the \vlsr\ of the sources.}
   \label{SiO-spectra}
   \end{figure}

\section{Observations\label{obs}}

Observations of the SiO isotopologue lines toward VY CMa, $o$ Ceti, W Hya, and R Leo were carried out with the 12-meter APEX telescope in 2011 September and 2012 December on Llano de Chajnantor in Chile. The single-sideband heterodyne receivers APEX-1 and APEX-2 \citep{Vassilev2008,Risacher2006} were used during the observations. The focus and pointing of the antenna were checked on Jupiter and Mars. The pointing and tracking accuracy were about 2\arcsec\ and 1\arcsec, respectively. The extended bandwidth Fast Fourier Transform Spectrometer \citep[XFFTS;][]{Klein2012} backend was mounted and configured into a bandwidth of 2.5 GHz and $\sim$ 0.1 \kms\ resolution. In addition, the {\it Herschel}/HIFI data of VY CMa, $o$ Ceti, W Hya, $\chi$ Cyg, R Cas, and R Dor were retrieved from the {\it Herschel} science archive. 


All spectra were converted to the main beam brightness temperature unit, $T_{\rm MB}$=$T_{\rm A}^{*}/\eta_{\rm MB}$ ($\eta_{\rm MB}$=$B_{\rm eff}/F_{\rm eff}$), using the forward efficiencies ($F_{\rm eff}$) and the beam-coupling efficiencies ($B_{\rm eff}$) from the APEX documentation\footnote{http://www.apex-telescope.org}. The beam efficiencies of HIFI were taken from the {\it Herschel}/HIFI documentation webpage. We adopted $\eta_{\rm MB}$ of 0.75, 0.73, and 0.76 for the APEX-1, 2, and HIFI data, respectively. All data were reduced and analyzed by using the standard procedures in the GILDAS\footnote{http://www.iram.fr/IRAMFR/GILDAS/} package. The SiO spectroscopic data were taken from the Cologne database for molecular spectroscopy (CDMS\footnote{http://www.astro.uni-koeln.de/cdms/}) and are listed in Table \ref{table-1}.

\section{Results and discussion \label{results}}

The APEX and {\it Herschel} SiO isotopologue spectra of the selected evolved stars (with both APEX and {\it Herschel} detections) are shown in Figures \ref{SiO-spectra} and \ref{SiO-spectra-2}, and the SiO intensity measurements are summarized in Table \ref{table-3} in the appendix. The \sioII\ and \sioIII\ emission is expected to be optically thin because the abundance of the main isotopologue \sioI\ is at least ten times larger than \sioII\ in the ISM \citep{Penzias1981}. Additionally, the solar and terrestrial \siII/\siIII\ ratios are close to 1.5 \citep{de Bievre1984,Anders1989}. The \sioII/\sioIII\ $J$=26--25 intensity ratios observed with the {\it Herschel}/HIFI instrument for $o$ Ceti and W Hya (Fig. \ref{SiO-spectra}) are consistent with the low-$J$ results obtained with the APEX telescope. Fitting two Gaussian profiles to the \sioII\ line and the partially blended \THCO\ $J$=10--9 line in the HIFI VY CMa spectra, we obtained a \sioII/\sioIII\ $J$=26--25 intensity ratio of 1.4$\pm$0.1, also consistent with the low-$J$ APEX data. Because the upper-state energies \Eup\ of $J$=26--25 lines are about 700 K higher than those of $J$=6--5, the constant \sioII/\sioIII\ intensity ratio of low- and high-$J$ transitions indicates optically thin \sioII\ and \sioIII\ emission with similar distributions and excitation conditions. {In addition, we believe that the \sioII\ and \sioIII\ emission obtained for our sample stars is unlikely to be dominated by masing effects due to the lack of any narrow spectral features.} Therefore, the \sioII/\sioIII\ intensity ratio directly reflects the abundance ratio between \siII\ and \siIII\ in the circumstellar envelopes of these stars, assuming any differences in chemical fractionation or photodissociation are minor. The derived \siII/\siIII\ ratios are listed in Table \ref{table-2} in the appendix.


\subsection*{Silicon isotope ratios}

Since \siI\ is mainly produced via the $\alpha$-process in massive stars, the \siI\ in low-mass stars comes from their natal clouds. Additionally, stable isotopes \siII\ and \siIII\ can be formed via slow neutron capture (the {\it s}-process) in both low- and high-mass stars. It has been shown by \citet{Timmes1996} that \siI\ is the primary isotope in the GCE with a roughly constant silicon-to-iron ratio over time, independent of the initial metallicity. On the other hand, neutron-rich isotopes \siII\ and \siIII\ show strong dependence on the composition and initial metallicity.



\begin{figure}
   \centering
   \includegraphics[angle=0,width=0.48\textwidth]{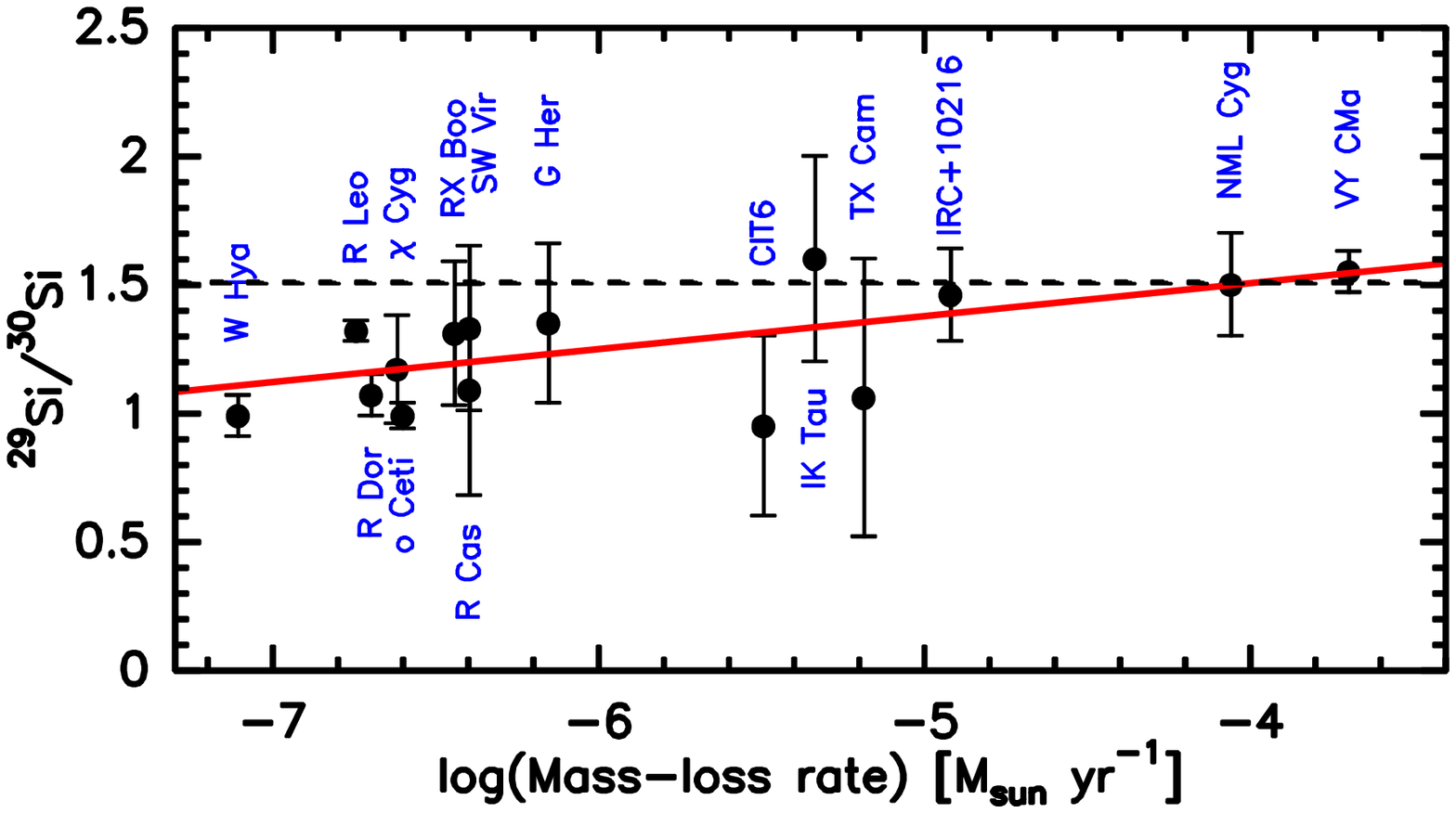}
   \caption{Comparison of $^{29}$Si/$^{30}$Si in evolved stars. The dashed line indicates the terrestrial and solar $^{29}$Si/$^{30}$Si abundance ratio of 1.51 \citep{de Bievre1984,Anders1989,Asplund2009}. The red line is a linear fit to the \siII/\siIII-$\dot{M}$ relation.}
   \label{Si-ratio}
   \end{figure}

In Figure \ref{Si-ratio}, the \siII/\siIII\ ratios derived from the SiO integrated intensities are plotted against the mass-loss rates for different evolved stars, and they show a tendency to increase with increasing mass-loss rates. The two supergiants VY CMa and NML Cyg and the carbon star IRC+10216 have \siII/\siIII\ ratios close to the solar value of 1.5. The rest of the samples (see also Table \ref{table-2}) have \siII/\siIII\ ratios $<1.5$, for example, the \siII/\siIII\ $\approx1$ for W Hya. There are two possibilities for the different \siII/\siIII\ ratios seen in our sample. One is that the silicon isotope ratios merely reflect the initial chemical composition of the environment where these stars were born and the different ratios are the results of different ages, which mainly depend on their masses and metallicities. The other possibility is that the stellar evolution can significantly change the silicon isotope ratios.

%
%


The first possibility implies that the \siII/\siIII\ ratio in the ISM has not significantly changed in the past 4.6 Gyr when the Sun was born. In comparison, VY CMa and IRC+10216 were born $\sim10^7$ and $1-5\times10^8$ years ago, assuming masses of 25--32 and 3--5 \Msol, respectively \citep[see][]{Portinari1998}. We found stars that we believe to be significantly
older than the Sun, such as W Hya (based on the mass-loss rate, initial mass, and current age), to have lower \siII/\siIII\ ratios. For instance, with an initial mass of 1--1.2 \Msol, W Hya has an age of 5--10 Gyr. In either the lower or higher age limit, this suggests a significant change in the \siII/\siIII\ ratio between the pre- and post-solar period: the \siII/\siIII\ ratios in the ISM increase from about 1 to 1.5 between 5 to 10 Gyr ago and remain roughly constant after the Sun was born. Given the time it takes low-mass stars to evolve onto the AGBs, it is unlikely that many low-mass AGB stars existed in our Galaxy between 5 and 10 Gyr ago, even if they had been formed at the beginning of the Milky Way formation. It is therefore also unlikely that low-mass AGB stars were significant contributors to the GCE in the presolar era. The \siII/\siIII\ ratio in the presolar era may be due to supernovae and/or other massive evolved stars. We note that the stars in our sample only trace the \siII/\siIII\ ratio of their natal clouds if they do not modify this ratio via nucleosynthesis \citep[see, e.g.,][]{Zinner2006}.




The second possibility for different \siII/\siIII\ ratios is that the stars in the AGB phase can significantly modify these ratios. Some of the M-type stars will become C-type stars after several dredge-up episodes with higher mass-loss rates toward the end of the AGB phase \citep[see, e.g.,][]{Herwig2005}. If the \siII/\siIII\ ratio can be modified by the $s$-process in the He-burning shell in evolved stars, it must be done efficiently because the AGB time scale is short \citep[a few times 10$^6$ yr, see][]{Marigo2007}. However, the modeling results of \citet{Zinner2006} show that the \siII/\siIII\ ratios of low-mass stars do not significantly change during the AGB phase \citep[see also the discussion of][]{Decin2010}.




\subsection*{\siII/\siIII\ ratio in presolar grains}

   \begin{figure*}
   \centering
   \includegraphics[angle=0,width=0.71\textwidth]{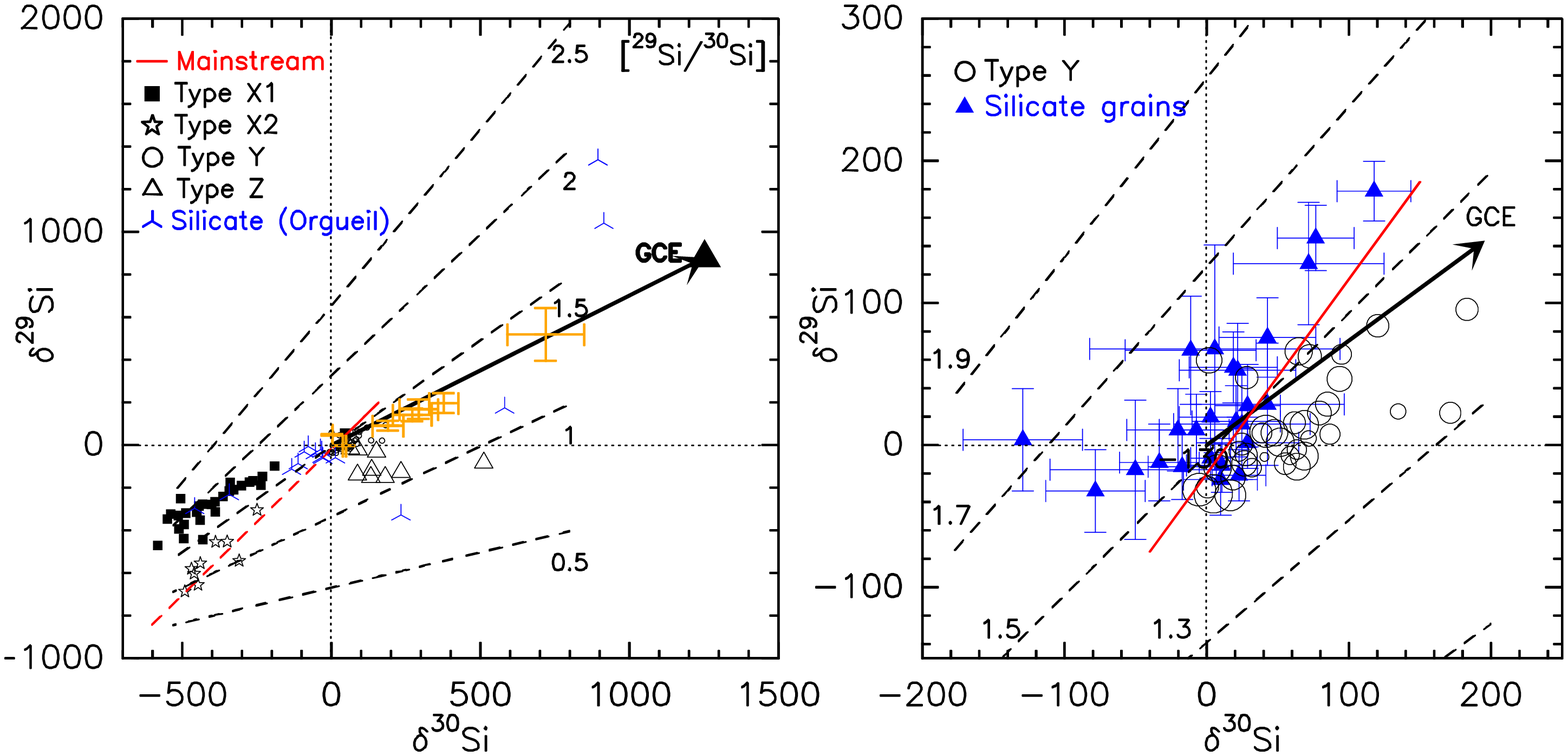}
   \caption{Silicon three-isotope plot for presolar grains. The delta notation is defined as $\delta^i{\rm Si}/^{28}{\rm Si}=[(^i{\rm Si}/^{28}{\rm Si})/(^i{\rm Si}/^{28}{\rm Si})_{\odot}-1]\times1000$. The \dsiII/\dsiIII\ ratios do not translate directly into the \siII/\siIII\ ratios, which are plotted as black dashed lines. Left: The data of different subgroups (X, Y, and Z) of SiC grains were taken from \citet{Lin2002}, \citet{Amari2001}, and \citet{Hoppe1997}, and the Orgueil silicate grains from \cite{Zinner2013}. The mainstream type ($\sim$93\%) of grains are indicated as a solid red line with a slope of 1.37. The dashed red line indicates the possible extension from the mainstream grains to X2 grains. The filled triangle indicates the ISM value \citep{Wolff1980}, and the black arrow indicates the direction of the Galactic chemical evolution. The orange crosses are the evolved star sample (about 3 \Msol) from \citet{Tsuji1994}. Right: Similar plot as the left one, but on a smaller scale, and the different silicate grain data were taken from \citet{Nguyen2007}, \citet{Mostefaoui2004}, and \citet{Nagashima2004}.
   }
   \label{Si-ratio-grain}
   \end{figure*}

Assuming the \siII/\siIII\ ratio in the gas-phase SiO is the same as it condenses onto dust grains or forms silicates, this primitive \siII/\siIII\ ratio may be carried by those grains when they are incorporated into new stellar and planetary systems. The \siII/\siIII\ ratio in presolar SiC grains has been studied in some meteorites \citep[e.g., the Murchison meteorite, see the review by][]{Zinner1998}. They have been categorized into different types (e.g., X, Y, and Z) according to their silicon isotopic anomalies. Most of the SiC grains found in meteorites are the so-called mainstream grains ($\sim$93\%, see Fig. \ref{Si-ratio-grain}), i.e., those with a slope of 1.34 on a silicon three-isotope plot \citep{Hoppe1994}. On the other hand, SiO is expected to condense onto the dust formation regions near O-rich stars, or via a possible heteromolecular nucleation of Mg, SiO, and \water\ to form silicates \citep{Goumans2012}. In the studies of Si isotopes in primitive silicate grains, the Si isotopic compositions of the majority of presolar silicates are similar to the SiC mainstream grains \citep{Nguyen2007,Mostefaoui2004,Nagashima2004,Vollmer2008}, indicating that the amount of Si isotopes locked in the SiC grains and the -SiO group in silicates may be similar; an example are the Orgueil silicate grains shown in Figure \ref{Si-ratio-grain}.



Most of the presolar grains have \siII/\siIII\ ratios around 1.5, but evidence of lower \siII/\siIII\ ratios are also found in presolar SiC grains, for instance, types X2 and Z in Figure \ref{Si-ratio-grain}. The type Z grains may have originated from a nearby evolved star \citep[see also][]{Zinner2006}. Additionally, the type X grains have been proposed to have a supernova origin \citep[e.g.,][]{Amari1992,Hoppe1994}, and have two or more subgroups \citep[see, e.g.,][]{Hoppe1995,Lin2002} with possible different stellar origins. According to the study of \citet{Lin2002} on the Qingzhen enstatite chondrite, the subgroups X1 and X2 show somewhat similar N and O isotopes abundance ratios, but have different slopes on the Si three-isotope plot (0.7 vs. 1.3 for X1 and X2, respectively). Because the metallicity in the local ISM is increasing owing to the GCE, $\delta^{29}$Si and $\delta^{30}$Si will increase with time accordingly. It is possible that the lower \siII/\siIII\ ratios seen in X2 grains may have originated from a population of evolved stars (such as the evolved stars with lower \siII/\siIII\ ratios). On the other hand, X1 grains with higher \siII/\siIII\ ratios are likely to be attributable to Type II supernovae \citep[see also][]{Zinner2013}. Moreover, it is important to point out that the higher-mass (about 3 \Msol) evolved star sample from \citet{Tsuji1994} can be well explained by the GCE (Fig. \ref{Si-ratio-grain}), considering the possible uncertainty in the \siII/\siIII\ ratio estimate for the present-day ISM.

\section{Conclusions}

We investigated the \siII/\siIII\ ratios of 15 evolved stars from the thermal SiO isotopologue emission obtained by the APEX and {\it Herschel} telescopes and from the literature. The inferred \siII/\siIII\ ratios tend to be lower among the older low-mass O-rich stars. Because the \siII/\siIII\ ratios are not significantly modified during the AGB phase and the contributions from the low-mass AGB stars are less important due to their long lifetimes, the lower \siII/\siIII\ ratios imply different enrichment of \siII\ and \siIII\ in the Galaxy between 5 to 10 Gyr ago with a nearly constant value of 1.5 after that. Noting that presolar grains may also have \siII/\siIII\ ratios lower than 1.5 (i.e., Type X2 and Z), we suggest that these grains could have been produced by one or more AGB stars with masses high enough to evolve onto the AGB in time to contribute to presolar grains. 




\begin{acknowledgements}
We thank the Swedish APEX staff for preparing observations and the referee for helpful comments. MGR gratefully acknowledges support from the National Radio Astronomy Observatory (NRAO). The National Radio Astronomy Observatory is a facility of the National Science Foundation operated under cooperative agreement by Associated Universities, Inc. IdG acknowledges the Spanish MINECO grant AYA2011-30228-C03-01 (co-funded with FEDER fund).
\end{acknowledgements}

\Online

   \begin{figure*}
   \centering
   \includegraphics[angle=0,width=0.8\textwidth]{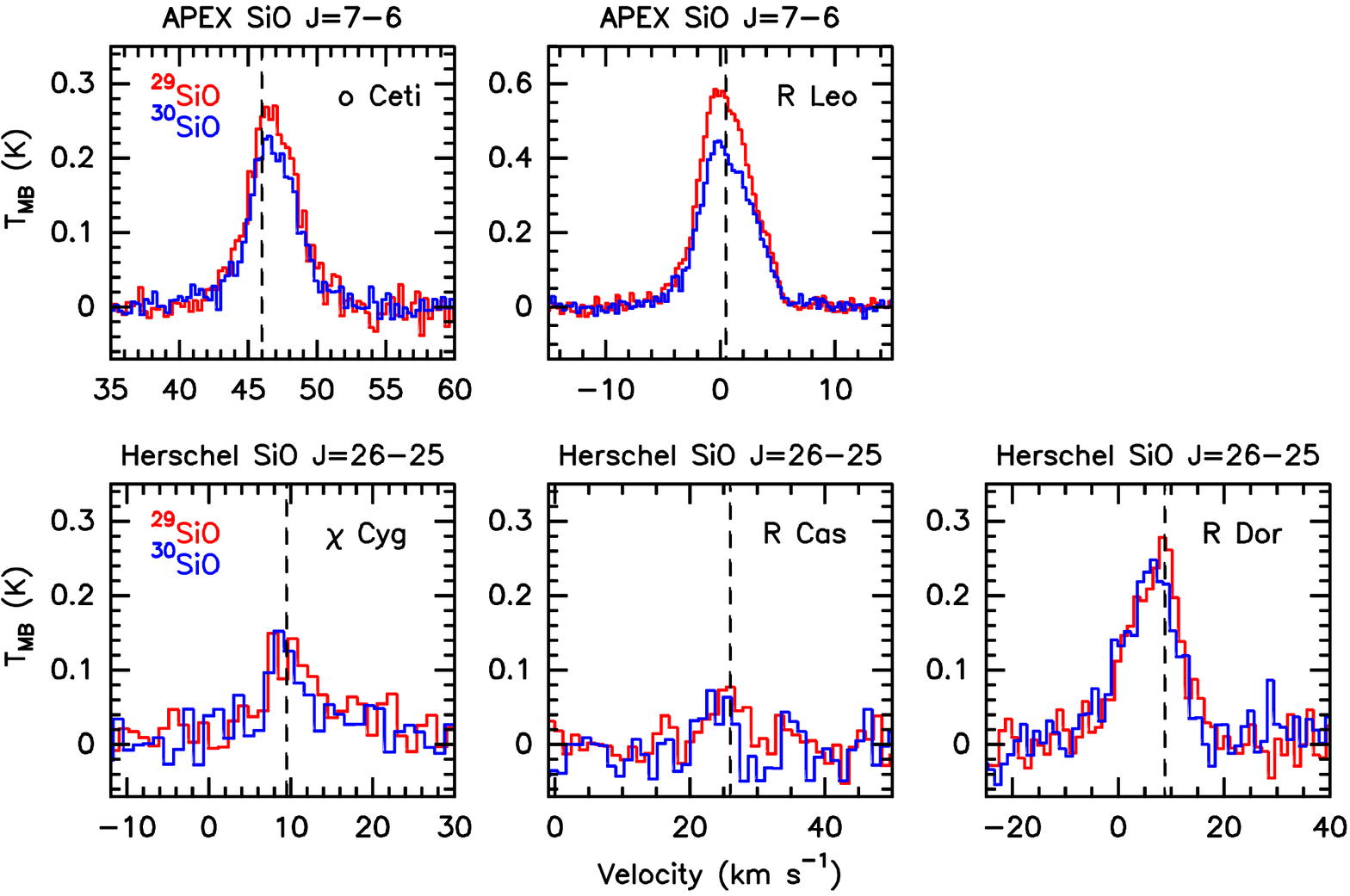}
   \caption{Upper panels: APEX ground-vibrational \sioII\ (red) and \sioIII\ (blue) $J$=7--6 spectra toward $o$ Ceti and R Leo. Lower panels: $Herschel$/HIFI ground-vibrational \sioII\ (red) and \sioIII\ (blue) $J$=26--25 spectra at around 1.1 THz toward $\chi$ Cyg, R Cas, and R Dor. The dashed lines indicate the \vlsr\ of the sources.}
   \label{SiO-spectra-2}
   \end{figure*}

\begin{table*}
\caption{APEX and {\it Herschel} SiO integrated intensity measurements}             
\label{table-3}      
\centering                          
\begin{tabular}{lcccccccc}        
\hline\hline                 
Source     &   \sioI\ 6--5   &  \sioII\ 6--5 & \sioIII\ 6--5  &  \sioII\ 7--6 & \sioIII\ 7--6   & \sioII\ 26--25   & \sioIII\ 26--25 \\    
           &   (K \kms)      &  (K \kms)     & (K \kms)      &  (K \kms)     &  (K \kms)       &  (K \kms)       &  (K \kms)       \\
       
\hline

VY CMa     & 107.4$\pm$0.3 &  24.8$\pm$0.3 & 16.3$\pm$0.3  & ...            & ...           & 10.1$\pm$0.7  & 6.4$\pm$0.3 \\
R Cas      &   ...           &   ...           &  ...           & ...            & ...            &  0.4$\pm$0.1 & 0.3$\pm$0.1 \\

$o$ Ceti   &   2.3$\pm$0.1 &  1.0$\pm$0.1  &  0.8$\pm$0.1 & 1.2$\pm$0.1  & 1.0$\pm$0.1  &  0.9$\pm$0.1 & 1.2$\pm$0.1 \\
$\chi$ Cyg &   ...           &   ...           &  ...           & ...            & ...            &  0.9$\pm$0.1 & 0.8$\pm$0.1 \\
R Dor      &   ...           &   ...           &  ...           & ...            & ...            &  3.2$\pm$0.2 & 3.0$\pm$0.2 \\
R Leo      &   ...           &   ...           &  ...           & 3.3$\pm$0.1  & 2.5$\pm$0.1  &  ...           & ...           \\
W Hya      & 18.4$\pm$0.1  &  4.8$\pm$0.1  &  4.3$\pm$0.1 & ...            & ...            & 1.1$\pm$0.1  & 1.2$\pm$0.1 \\

\hline                                   
\end{tabular}

\tablefoot{The integrated intensities are measured in $T_{\rm MB}$ and do not include the calibration errors of the APEX and {\it Herschel} telescopes (from a few percent up to 10\%) because the \sioII\ and \sioIII\ lines were detected at the same band simultaneously and their ratios do not strongly affected by the calibrations.
}

\end{table*}

\begin{table*}
\caption{Overview of envelope terminal velocities, mass-loss rates, and \siII/\siIII\ ratios toward the selected evolved stars}             
\label{table-2}      
\centering                          
\begin{tabular}{lccccccc}        
\hline\hline                 
Source     &   $d$   &  $V_{\rm e}$ & $\dot{M}$           &  \siII/\siIII\tablefootmark{a}  & Spectral Type & Stellar Type  & Note \\    
           &   (pc)  &  (\kms)      & (\Msol\ yr$^{-1}$)  &                &               &         &      \\
       
\hline

VY CMa     &  1170 &  46.5  &  $2.0\times10^{-4}$ & $1.55\pm0.08$ & M2/4II    & RSG          & APEX-1+HIFI\tablefootmark{d} \\
NML Cyg    &  1610 &  33.0  &  $8.7\times10^{-5}$ & $1.50\pm0.20$ & M6I       & RSG          & \citet{Tsuji1994}\\ 
IRC+10216  &   120 &  14.5  &  $1.2\times10^{-5}$ & $1.46\pm0.18$ & C9,5e     & MIRA         & \citet{Tsuji1994} \\
IK Tau     &   260 &  18.5  &  $4.6\times10^{-6}$ & $1.60\pm0.30$\tablefootmark{b} & M8/10IIe  & MIRA    & \citet{Decin2010,Kim2010} \\
TX Cam     &   380 &  21.2  &  $6.5\times10^{-6}$ & $1.06\pm0.54$ & M8.5      & MIRA         & \citet{Cho1998} \\
CIT 6       &   440 &  20.8  &  $3.2\times10^{-6}$ & $0.95\pm0.35$ & Ce        & SRa          & \citet{Zhang2009} \\
G Her      &   310 &  13.0  &  $7.0\times10^{-7}$ & $1.35\pm0.31$ & M6III     & SRb          & \citet{Tsuji1994} \\
R Cas      &   106 &  13.5  &  $4.0\times10^{-7}$ & $1.09\pm0.41$ & M7IIIe    & MIRA         & HIFI\tablefootmark{d} \\
SW Vir     &   170 &   7.5  &  $4.0\times10^{-7}$ & $1.33\pm0.32$ & M7III     & SRb          & \citet{Tsuji1994}\\
RX Boo     &   155 &   9.0  &  $3.6\times10^{-7}$ & $1.31\pm0.28$ & M7.5e     & SRb          & \citet{Tsuji1994}\\
$o$ Ceti   &   107 &   8.1  &  $2.5\times10^{-7}$ & $1.04\pm0.06$ & M7e       & MIRA         & APEX-1/2+HIFI\tablefootmark{d} \\
$\chi$ Cyg &   149 &   8.5  &  $2.4\times10^{-7}$ & $1.06\pm0.22$\tablefootmark{c} & S6        & MIRA         & HIFI\tablefootmark{d} \\
R Dor      &    45 &   6.0  &  $2.0\times10^{-7}$ & $1.14\pm0.16$ & M8IIIe    & SRb          & HIFI\tablefootmark{d} \\
R Leo      &   130 &   5.0  &  $1.8\times10^{-7}$ & $1.32\pm0.04$ & M8IIIe    & MIRA         & APEX-2\tablefootmark{d} \\
W Hya      &    77 &   8.5  &  $7.8\times10^{-8}$ & $0.99\pm0.05$ & M7e       & SRa          & APEX-1+HIFI \tablefootmark{d}\\

\hline                                   
\end{tabular}

\tablefoot{The data of distance $d$, terminal velocity of CO envelope $V_{\rm e}$, and mean mass-loss rate $\dot{M}$ of the selected sources were compiled from \citet{Woods2003}, \citet{De Beck2010}, \citet{Justtanont2012}, and \citet{Schoier2013}. We note that \citet{Schoier2013} adopted smaller distances for $\chi$ Cyg (110 pc) and RX Boo (120 pc), and the distance to VY CMa was averaged from two measurements \citep{Choi2008,Zhang2012}. The mean \siII/\siIII\ ratios for VY CMa, $o$ Ceti, and W Hya were derived from the thermal \sioII/\sioIII\ emission ratios obtained by the APEX and {\it Herschel} telescopes. The \siII/\siIII\ ratios for R Cas, $\chi$ Cyg, and R Dor were derived from the {\it Herschel}/HIFI data. The \siII/\siIII\ ratios of G Her, SW Vir, and RX Boo were taken from \citet{Tsuji1994}, TX Cam from \citet{Cho1998}, and CIT 6 from \citet{Zhang2009}. Spectral types were take from the SIMBAD database, \citet{De Beck2010}, and the references therein. 
\tablefoottext{a}{The \siII/\siIII\ ratios are the mean values (equally weighted)} when more than one transition was detected.
\tablefoottext{b}{The {\it Herschel}/HIFI data only show a 2-$\sigma$ detection of \sioII\ and \sioIII\ $J$=26--25 lines with a ratio of 1.6$\pm$0.8. However, judging from the APEX $J$=8--7 and 7--6 data of \citet{Kim2010}, the integrated ratios are also close to 1.6. We note that the baselines in the \sioIII\ spectra of \citet{Kim2010} may be too low. Therefore, the actual \sioII/\sioIII\ ratio should be $\lesssim 1.6$ \citep[see also the discussion of][who adopted an abundance ratio of 3$\pm$2 from their modeling results]{Decin2010}.}
\tablefoottext{c}{The ratio reported by \citet{Ukita1988} and \citet{Tsuji1994} is about 2.4 with a large uncertainty (same as V1111 Oph) estimated from $J$=2--1, which may be affected by masing effects. Therefore, we estimated the \sioII/\sioIII\ ratio only from the new HIFI measurements.}
\tablefoottext{d}{The SiO intensity measurements are listed in Table \ref{table-3}.}
}

\end{table*}


\end{document}